# General Effect Modelling (GEM)

## Part 2. Multivariate GEM applied to gene expression data of type 2 diabetes detects information that is lost by univariate validation

Mosleth, E.F.[1,+], Dankel, S.N.[2,3,4], Mellgren, G.[2,3,4], Barajas-Olmos F.[5], Orozco, L.S.[5], Lysenko, A.[6], Ofstad, R.[1], Begum, M.C.[2], Martens, H.[7], and Liland, K.H.[8.]

[1] *Nofima, Norwegian Institute for Food, Fisheries and Aquaculture Research, Ås, Norway*
[2] *Mohn Research Center for Diabetes Precision Medicine, Department of Clinical Science, University of Bergen, Bergen, Norway*
[3] *Mohn Nutrition Research Laboratory, Department of Clinical Science, University of Bergen, Bergen, Norway.*
[4] *Hormone Laboratory, Haukeland University Hospital, Bergen, Norway.*
[5] *Immunogenomics and Metabolic Diseases Laboratory, National Institute of Genomic Medicine SS, Mexico City, Mexico.*
[6] *Laboratory for Medical Science Mathematics, RIKEN Center for Integrative Medical Sciences, Yokohama, Japan*
[7] *Idletechs AS, Trondheim, Norway*
[8] *Faculty of Science and Technology, Norwegian University of Life Sciences, 1430 Ås, Norway.*

[+] *Corresponding author:* ellen.mosleth@nofima.no

## Abstract

General Effect Modelling (GEM) is an umbrella over different methods that utilise effects in the analyses of data with multiple design variables and multivariate responses. To demonstrate the methodology, we here use GEM in gene expression data where we use GEM to combine data from different cohorts and apply multivariate analysis of the effects of the targeted disease across the cohorts. Omics data are by nature multivariate, yet univariate analysis is the dominating approach used for such data. A major challenge in omics data is that the number of features such as genes, proteins and metabolites are often very large, whereas the number of samples is limited. Furthermore, omics research aims to obtain results that are generically valid across different backgrounds. The present publication applies GEM to address these aspects. First, we emphasise the benefit of multivariate analysis for multivariate data. Then we illustrate the use of GEM to combine data from two different cohorts for multivariate analysis across the cohorts, and we highlight that multivariate analysis can detect information that is lost by univariate validation.

## Keywords
Multivariate analysis, General Effect Modelling (GEM), omics data, Type 2 diabetes

# Introduction

We are now in a situation where very large data can be obtained in many different fields. The obtained response is often a result of several factors. General Effect Modelling (GEM) is a methodology for experiments with multiple design variables and multivariate responses. GEM can also be used for collected data, where several variables can be considered to influence the obtained data. GEM is developed as a flexible umbrella that covers different data analytical communities for handling multivariate data with multiple input design variables and multivariate responses.

In the present publication, we use GEM on omics data for the demonstration of the methodology. Omics data are comprehensive data. The analysis of all genes present in an organism is called genomics, and the path from the genes to the final phenotype is a chain of omics features (**Figure 1**). Transcriptome is the expression levels of mRNA molecules, which is a mirror copy of one gene that is activated. The proteome is the proteins, the metabolome is the metabolites, and the phenome is the final phenotypic expression. There are also other omics, such as epigenomics, lipidomics etc. Although there are technical limitations to how many features within each category can be observed, the omics data in general contain very many variables and often few samples relative to the number of variables. This results in the typical wide and short data tables.

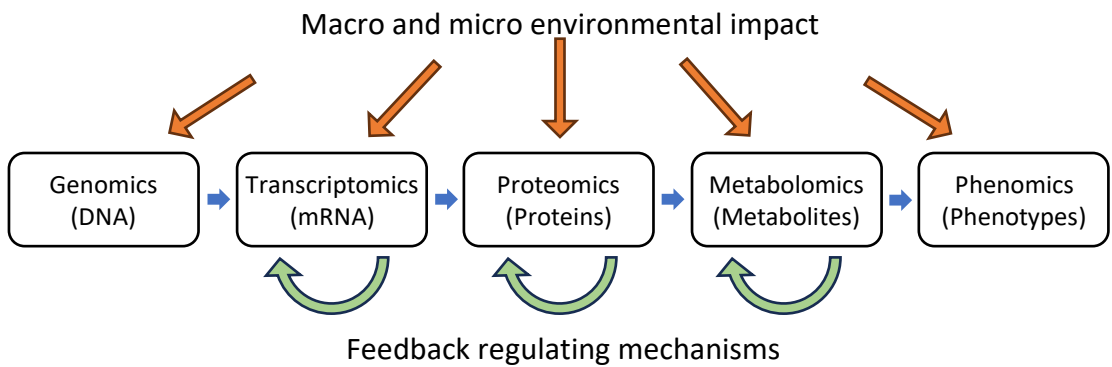

**Figure 1**. *Overview of the flow of data in functional omics from the genome (deoxyribonucleic acid, DNA), transcriptome (gene expression data, the mRNA messenger), proteome, metabolome to the final phenotypic expression, which may be observed as quality, disease pattern etc*.

The processes in the cells are extremely well controlled, and the modern analytical methods of omics data are detailed and often highly precise. This enables insight into cellular processes and underlying causes on a level that was never possible before. Due to the complexity of biological processes, there is a need for data analytical tools that can handle complexity.

Importantly, the path from the genome to the final phenotypic expression is not a linear line of one-to-one connections. For example, there are many more proteins than genes, and regulation mechanisms at all steps. A phenotypic character can be controlled at different levels, for example: by the composition of the genes, by epigenetics, by transcriptome factors, on the protein level or on the metabolome levels. The causality in the functional omics chain can therefore better be characterized by many-to-many relationships [1]. Therefore, it is highly relevant to take a multivariate approach in omics studies where multiple genes are considered simultaneously to shed light on the topic under consideration.

Yet, univariate analyses are still dominating omics research. A variant of the t-test is often used as a statistical test to compare two groups, and the experimental design is often limited to comparing two groups as that can be analysed by a t-test. This strongly limits the knowledge that can be achieved from omics research. Univariate analyses were developed for situations where data consisted of few features relative to the number of samples. The risk of false positives can be addressed by adjustment of the p-values for multiple comparisons and lists of differentially expressed features (genes, proteins etc.) can be loaded into enrichment analysis to consider the combined action of multiple features. However, this does not solve the initial problem of applying univariate statistical tests that consider only one feature at a time.

Multivariate analysis enables insight into the combined effects of multiple genes, proteins etc. In the present publication, we pay attention to two different families of multivariate analyses. Chemometrics [1-3] which uses bilinear modelling and projection onto latent space as the core methodology and Regularization methods [4-7].

*Bilinear modelling*

The most familiar method of bilinear modelling is Principal Component Analysis (PCA) [1, 2]. Partial Least Squares (PLS) can be considered as a two- (or more-) blocks extension of PCA [1, 3, 8]. PLS has also been renamed as Projection onto Latent Space (refr). Whereas each component in PCA is estimated to maximise the variance in the data, by PLS the covariance between data blocks is maximised under the estimation of the components. The components in PLS may be called PLS factors or just components.

In PCA and PLS, the estimated components are linear both with respect to the samples and with respect to the variables, thereby the name bilinear methods. A frequently asked question to bilinear methods is how these methods handle non-linearity as PCA and PLS are based on linear projections. The bilinear model is often found to work well for non-linear data. This can be illustrated by modelling "banana-shaped" data (**Figure 2**). The first components in PCA will go through the length direction of the banana, and the next component will find the orthogonal direction. The third component (not visualised in the figure) will capture the thickness of the banana. Thus, data formed as a banana can well be modelled by the bilinear methods.

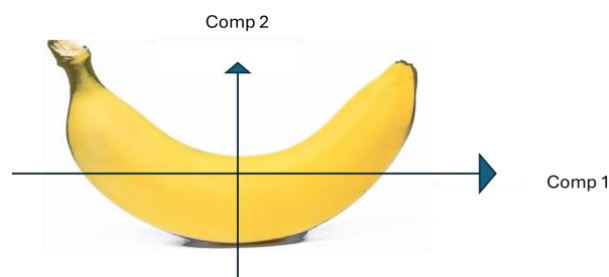

**Figure 2**. *Illustration of how PCA would model non-linear data such as the bent shape of a banana.*

Our visualisation of dimensions is limited to three as there are three dimensions in our physical room. However, in a data table of experimental results, any new set of variables that are independent of each other will form new dimensions. Most often, and always in omics data, there is strong interdependency between the variables. For example, many genes and proteins are regulated by common mechanisms, genes may be closely located on the chromosome, and genes may be controlled by common transcriptional control mechanisms. The true independent dimension in omics data is always small compared with the very large number of observed genes, proteins etc. The family of methods based on projection onto latent spaces search for the underlying dimension of the data that may reflect the underlying mechanisms of interest for the task under study.

*Regularization*

Regularization is a technique in machine learning that is used to prevent overfitting (overfitting) of the model to the training data. Regularization introduces a form of constraint on the model complexity by adding an extra term to the cost function during training. The goal is to reduce the weights of some features so that the model becomes less sensitive to small changes in the training data. Whereas projections on latent subspaces, as described above for PCA and PLS, **utilize** the multicollinearity in the data, the regularization technique is a strategy to **combat** multicollinearity.

## GEM

A mathematical description of GEM is presented in **Figure 3**, and for more details, we refer to Part 1 of this publication series.

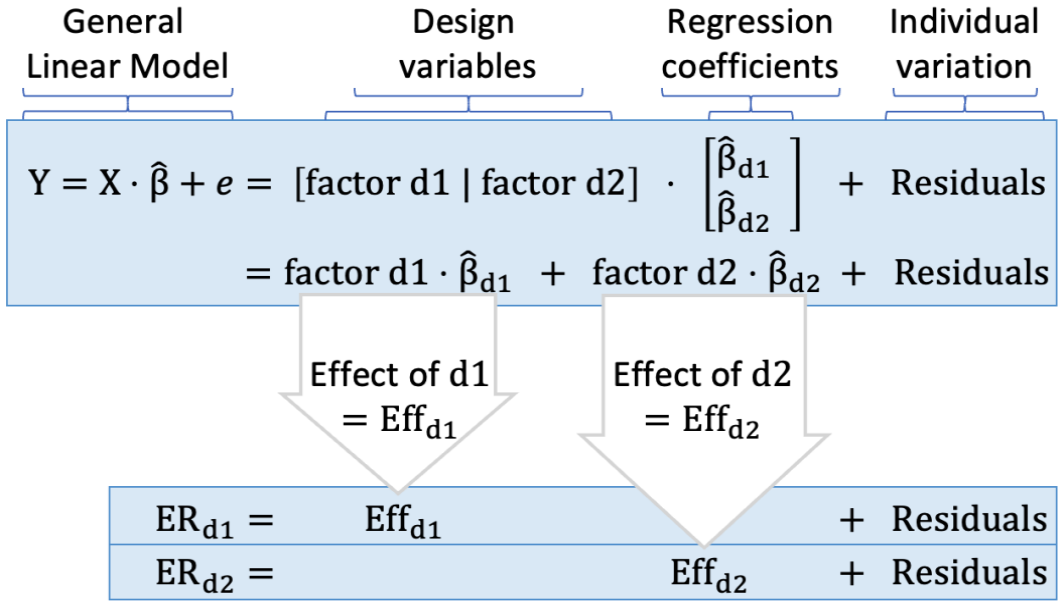

*Figure 3. Mathematical description of GEM shown for a study of two design variables (d1 and d2). The first step of GEM is the estimation of the effects of the design variables using a General Linear Model (GLM). X is the design matrix where d1 and d2 are columns. Each design variable is on two levels. The regression coefficients ($\hat{\beta}$'s) are estimated from the model, and the residuals contain the variation that can not be explained by the two design factors. The effects of one design variable at a time along with the residuals of the original GLM model are used in step 2 of GEM which can be any multivariate or univariate statistical analysis of the ER values. Mathematically described, the effects of one design variable at a time are included in the statistical analyses, whereas the effects of the other design variables are orthogonalized and omitted in the statistical analyses. The degrees of freedom consumed in the first step are taken into consideration in the following analysis.*

# GEM used to combine transcriptome data of type 2 diabetes from different cohorts

GEM is here used to combine data from different cohorts to increase the power of the statistical analysis and to identify differences that are valid across cohorts.
The case under study is to identification of gene expression levels altered by T2D that are not confounded by different levels of obesity. Confounding obesity and T2D is a challenge as the majority of persons with T2D also have obesity, whereas less than 15% of persons with obesity develop T2D. The data that are analysed are, therefore, balanced for obesity by including only persons with obesity challenges subjected to fat-reducing surgery.

## Material and method

Previously published gene expression data of subcutaneous adipose tissues from two independent cohorts is reanalysed to identify the expression levels altered by T2D. All persons had high BMI, and they were subjected to abdominal fat-reducing surgery. The first cohort ('cohort 1') is a Norwegian-Caucasian cohort [9, 10] and the second cohort ('cohort 2') is a Mexican cohort [11, 12]. Cohort 1 consisted of 8 T2D and 7 without T2D (nonD), and cohort 2 consisted of 14 T2D and 13 nonD. This gave a total of 42 persons across the two cohorts.

### Pre-analysis

Data explorations were performed by PCA to view the structure of the data and to discover if there is a pattern of variation that is not linked to the experimental design. The PCA revealed batch effects in cohort 2.

### GEM – to omit batch effects in cohort 2

The batch effect, detected for cohort 2, was removed by GEM. The R-codes are as follows:

```r
# Organise data as data.frame:
   Me.DF.SAT  <- data.frame(
       # an indicator variable for batch ("batch1" - "batch2"),
       factorA  = Me.Design.SAT$A,
       # disease status ("nonD" - "T2D"),
       factorB  = Me.Design.SAT$B,
       Features = I(Me.Features.SAT))
# GEM Step 1 (GLM):
   gem           <- GEM(Me.Features ~ factorA * factorB, data = Me.DF.SAT)
   Me.er.values.B <- gem$ER.values$factorB
```

For cohort 2, further analysis used the ER values of disease status.

### GEM – to omit combined data of cohorts 1 and 2

Thereafter we applied GEM to combine the data from the two different cohorts, using the original data of cohort 1 and ER values after omitting batch effect by GEM for cohort 2.

Cohort identity was used as one design variable, T2D as the other design variable, and we also included the interaction between these design variables. The R-codes are as follows:

```r
# Get common association identities and common design variables from the two cohorts
   common.genes     <- intersect(colnames(Genes.Cohort.1),colnames(Genes.Cohort.2))
   dataset1         <- scale(Genes.Cohort.1[,common.genes])
```

```r
    dataset2            <- scale(Genes.Cohort.2[,common.genes])
  Genes.Cohort.1.and.2  <- scale(rbind(dataset1,dataset2))
  common.design.variables <- intersect(colnames(Design.Cohort.1),colnames(Design.Cohort.2))
    dataset1            <- scale(Design.Cohort.1[,common.design.variables])
    dataset2            <- scale(Design.Cohort.2[,common.design.variables])
  Design.Cohort.1.and.2 <- scale(rbind(dataset1,dataset2))
    cohort              <- Design.Cohort.1.and.2$cohort
    disease             <- Design.Cohort.1.and.2$disease
# Organise data as data.frame.
  T2D.data <- data.frame(
     transcriptome = I(transcriptome),
     cohort  = factor(cohort),
     disease = factor(disease))
# GEM Step 1 (GLM):
    library(gemR)
    T2D.gem  <- GEM(transcriptome ~ cohort + disease + cohort:disease, data = T2D.data)
```

Multivariate analysis by PLS and Elastic net in the framework of GEM "GEM-PLS" and "GEM-Elastic net", respectively. This means that these were performed on the ER values obtained by the first step in GEM and that degrees of freedom are adjusted for in PLS analyses. Separate analyses were performed on the ER values of cohort, the ER values of disease status and the ER values of the interaction term.

In PLS, multiple feature selection methods can be applied for the identification and interpretation of genes that contribute to the multivariate model. Here we used Jackknife adapted to bilinear modelling, where the stability of the regression coefficients is tested by a t-test across multiple models performed in the cross-validation routine where one segment (a set of samples) is left out at the time from the modelling and used for validation [13]. We used a rejection limit of 0.05. Notably, t-test in Jackknife adapted to bilinear modelling is a test of the multivariate model, which is distinctly different from a t-test of the individual features in a univariate model.

The elastic net model was tuned to search for the minimum number of genes that could classify nonD from T2D based on the gene expression data

*Univariate analysis*

As benchmarking, univariate analysis by ANOVA was performed within each cohort and across the two cohorts. The r program "ffmanova" Langsrud [14] was applied. Adjustment of the p-values was performed by rotation test [15], by Bonferroni [16] and by Benjamin Hochberg [17]. The R-codes were as follows:

```r
# GEM Step 2 (Multivariate analysis):
# ANOVA and p-values adjusted for multiple comparisons:
    library(ffmanova)
    R          <- T2D.data$transcriptome
    B          <- T2D.data$disease # status ("nonD" vs "T2D")
    my.data    <- data.frame(B,R)
    FF.res     <- ffmanova(R ~ B, data = my.data, stand=TRUE, nSim = 1000, verbose=TRUE)
    p.Raw      <- FF.res$pRaw[-1,]
    p.FDR.ff   <- FF.res$pAdjFDR[-1,];        # FDR adjusted p-values by rotation test
    p.FDR.bh   <- p.adjust(p.Raw, "BH")       # Benjamini, Hochberg corrected p-values
```

```
p.FDR.bonf <- p.adjust(p.Raw, "bonferroni") # Bonferroni corrected p-values
```

## Results

### *Pre-analysis*

PCA of cohort 2 revealed a strong pattern of variation that dominated the first component (**Figure 4**). This pattern accounted for as much as 56% of the total variation in the gene expression data in PCA. It was found that this pattern was due to different running batches in the molecular analysis. The batch effect was omitted by GEM. Further analyses are performed on ER values of the disease status where batch effects are omitted.

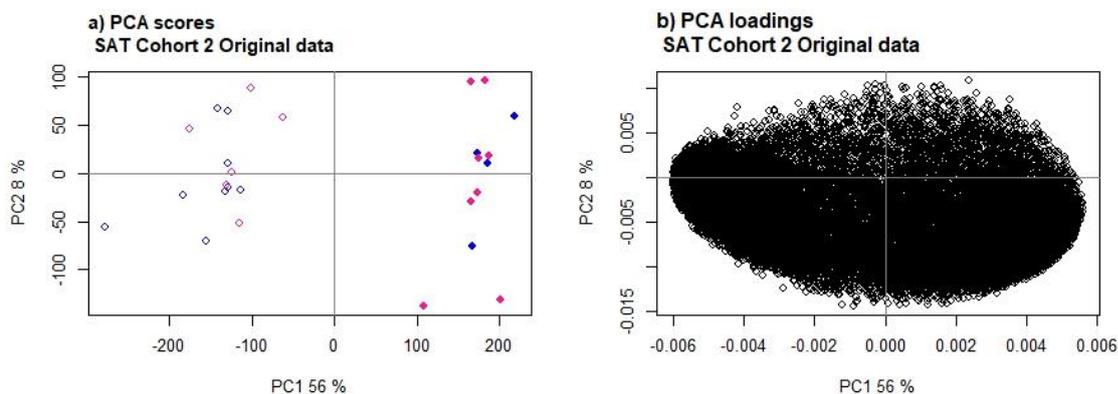

*Figure 4. Cohort 2. Explorative multivariate analysis by PCA. (**a**) Score plots of the persons displaying samples from one batch (open triangles) and the other batch (closed triangles) for T2D (pink) and nonD (blue). (**b**) Corresponding loading plot of the genes. The figure shows that a strong batch effect was found by the PCA analysis.*

### *Univariate analysis*

Univariate ANOVA applied within each cohort and across the cohorts after combining the cohorts into one larger data table using GEM are presented in **Figure 4**. Some of the genes had raw p-values below the most common rejection limits of 5%. However, when the p-values were adjusted for multiple comparisons, none of the genes displayed adjusted p-values close to the rejection limit.

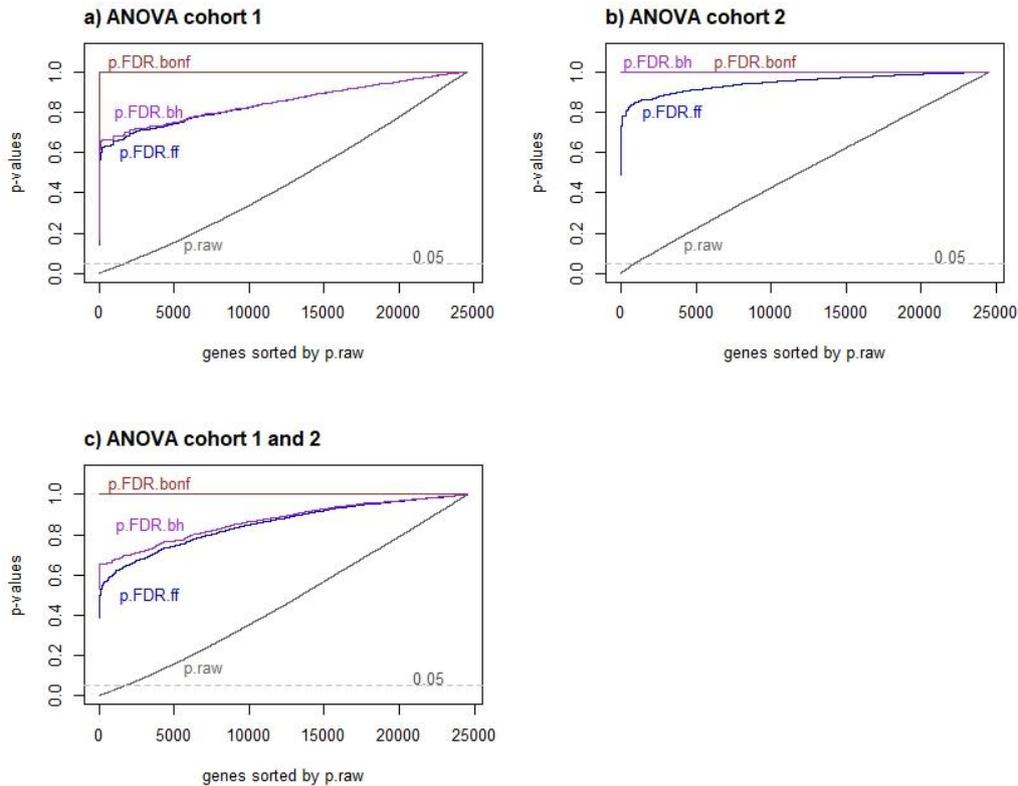

*Figure 5.* Cohort 1 and 2. Univariate ANOVA displayed as raw and p-values for multiple comparisons. **(a)** Cohort 1, **(b)** cohort 2, **(c)** cohort 1 and 2 combined by GEM. p.FDR.ff are p-values adjusted by rotation test, p.FDR.bh are p-values adjusted by Benjamin-Hochberg and p.FDR by Bonferroni corrected p-values.

*Multivariate GEM analysis*

GEM applied on the combined data of cohorts 1 and 2 were subjected to supervised multivariate analysis by GEM-PLS and GEM-Elastic net that identified genes for the separation of nonD from T2D. GEM-PLS analysis of each main effect and the interacting effects are displayed in **Figure 3**; GEM-PLS analysis on the effects of T2D (**Figure 3 a-c**), on the effects of cohorts (**Figure 3 d-f**), and the interacting effects between T2D and cohorts (**Figure 3 g-i**). In the GEM-PLS loading plot, only genes identified by Jackknife are displayed.

The score plot of the analysis of the disease status, T2D were separated from nonD by the two first components in GEM-PLS (**Figure 3 a**). Significant genes for T2D were identified both by Jackknife in GEM-PLS-DA (**Figure 3 b**) and by GEM-Elastic net (**Figure 3 c**).

The analysis of cohort (**Figure 3 d-f**) and interaction between T2D and cohort (**Figure 3 g-i**) did not result in any significant genes, which shows that there were no detectable main effects of cohort and the effect of T2D was consistent across the cohort.

Feature selection in a multivariate model has different aims than a p-value validation in a univariate model, as the multivariate model validates the combined effects of multiple genes. Furthermore, multivariate modelling and feature selection in Elastic net is different from multivariate modelling and feature selection in PLS.

The elastic net model was tuned to search for the minimum number of genes that could classify nonD from T2D based on the gene expression data. This resulted in 81 significant genes. A PLS model projects the main information in the data down to underlying latent variables that can be interpreted for the separation of nonD and T2D. According to Jackknife in GEM-PLS, 1806 genes were significant, which means that they have stable regression coefficients across the cross-validation routine. The genes identified by Jackknife in GEM-PLS included all the genes identified by GEM-Elastic net (except one gene: hook microtubule tethering protein 1, *HOOK1*).

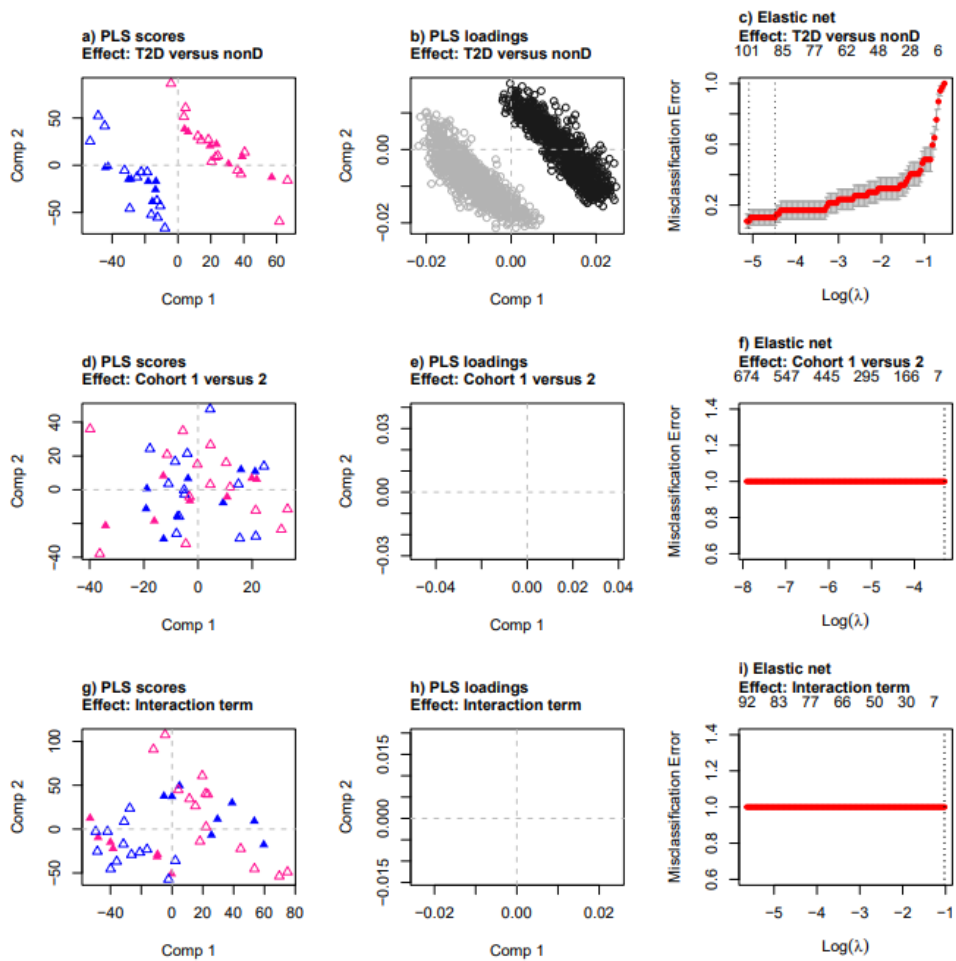

**Figure 5.** *Cohort 1 and 2. Supervised multivariate analysis by PLS-DA in GEM (GEM-PLS) where separate analyses were performed of the effects of (**a-c**) T2D, (**d-f**) cohort and (**g-i**) their interaction. (**a,d,g**) Score plots of the persons displaying **cohort 1** (open triangles) and **cohort 2** (closed triangles) for T2D (pink) and nonD (blue). (**b,e,h**) Corresponding loading plot of the genes where only genes significant by Martens uncertainly is displayed, which resulted in significant genes for the analysis of T2D compared with nonD (**b**), but no genes in the two latter analyses (**e,h**) thereby empty plots. (**c,f,i**) Elastic net which identified significant genes in the analysis of T2D compared with nonD (**b**), but no genes in the two latter analyses (**e,h**).*

Boxplots of some of the genes identified both by Elastic net and by Jackknife in PLS are displayed in **Figure 6** (downregulated genes) and **Figure 7** (upregulated genes).

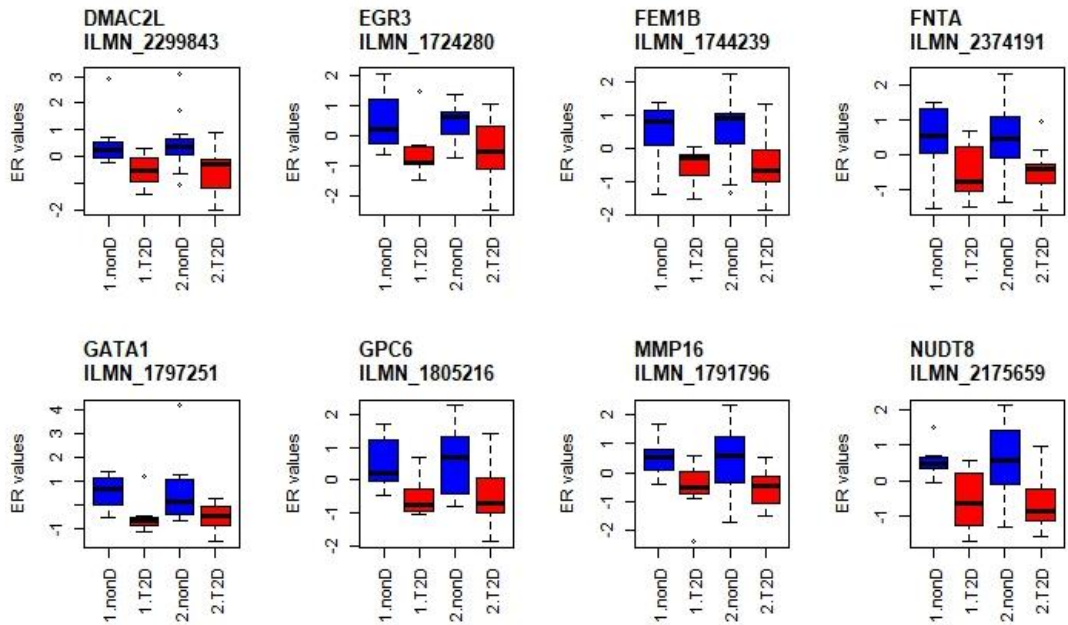

**Figure 6.** *Box plots of the expression levels of genes significantly downregulated for T2D by GEM-PLS and by GEM-Elastic net in cohort 1 (columns 1 and 2 to the left) and cohort 2 (columns 3 and 4 to the right) for nonD (blue) and T2D (red).*

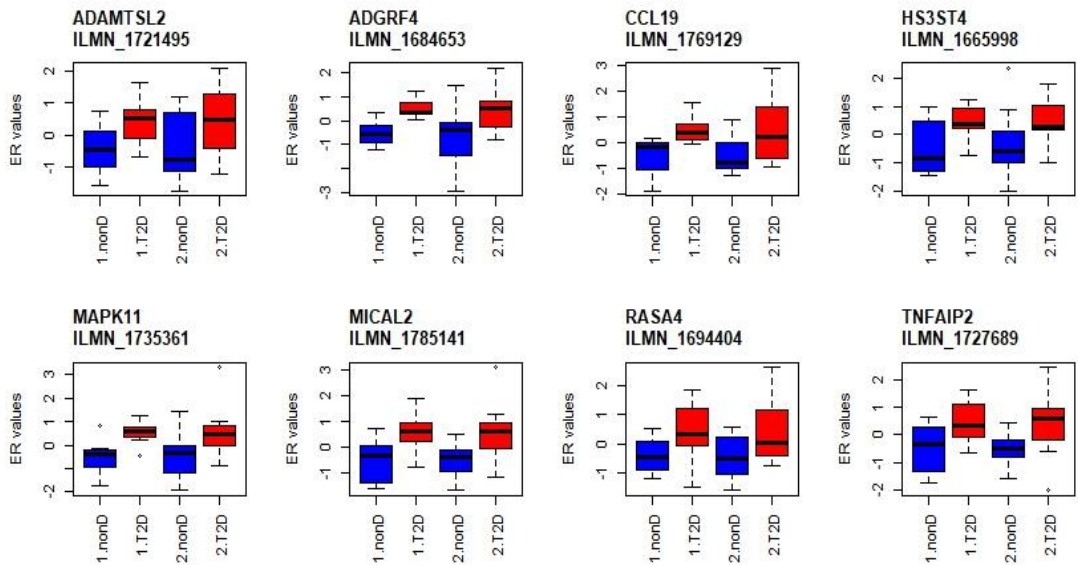

**Figure 7.** *Box plots of the expression levels of genes significantly upregulated by GEM-PLS and by GEM-Elastic net for T2D in cohort 1 (columns 1 and 2 to the left) and cohort 2 (columns 3 and 4 to the right) for nonD (blue) and T2D (red).*

As shown in the multivariate analysis in **Figure 5**, there is a need for multiple genes to separate nonD from T2D. Although the expression levels of two genes are not sufficient to separate nonD from T2D we display in **Figure 8**, 1D plots and 2D plots of two of the genes to illustrate a multivariate analysis. The gene [F-actin]-monooxygenase (*MICAL2*) was identified by both multivariate analyses as positively associated with T2D. The mitochondrial import receptor subunit TOM70 (*TOMM70*) was observed by Jackknife in GEM-PLS, but not by GEM-Elastic net.

As considered alone in a univariate test, *TOMM70* is not significantly lower for T2D than for nonD. A PLS model does, however, consider how *TOMM70* contributes along with the other genes. Although more than two genes are needed to separate nonD and T2D, the 2D plot of *TOMM70* and *MICAL2* in **Figure 8** illustrates that a combined view of the two genes contributes to a better separation of nonD versus T2D than any of these genes alone.

Details of the results of GEM-PLS for these cohorts will be presented elsewhere, where we also include yet another cohort of persons with and without T2D, balanced for BMI (Mosleth et al. in prep).

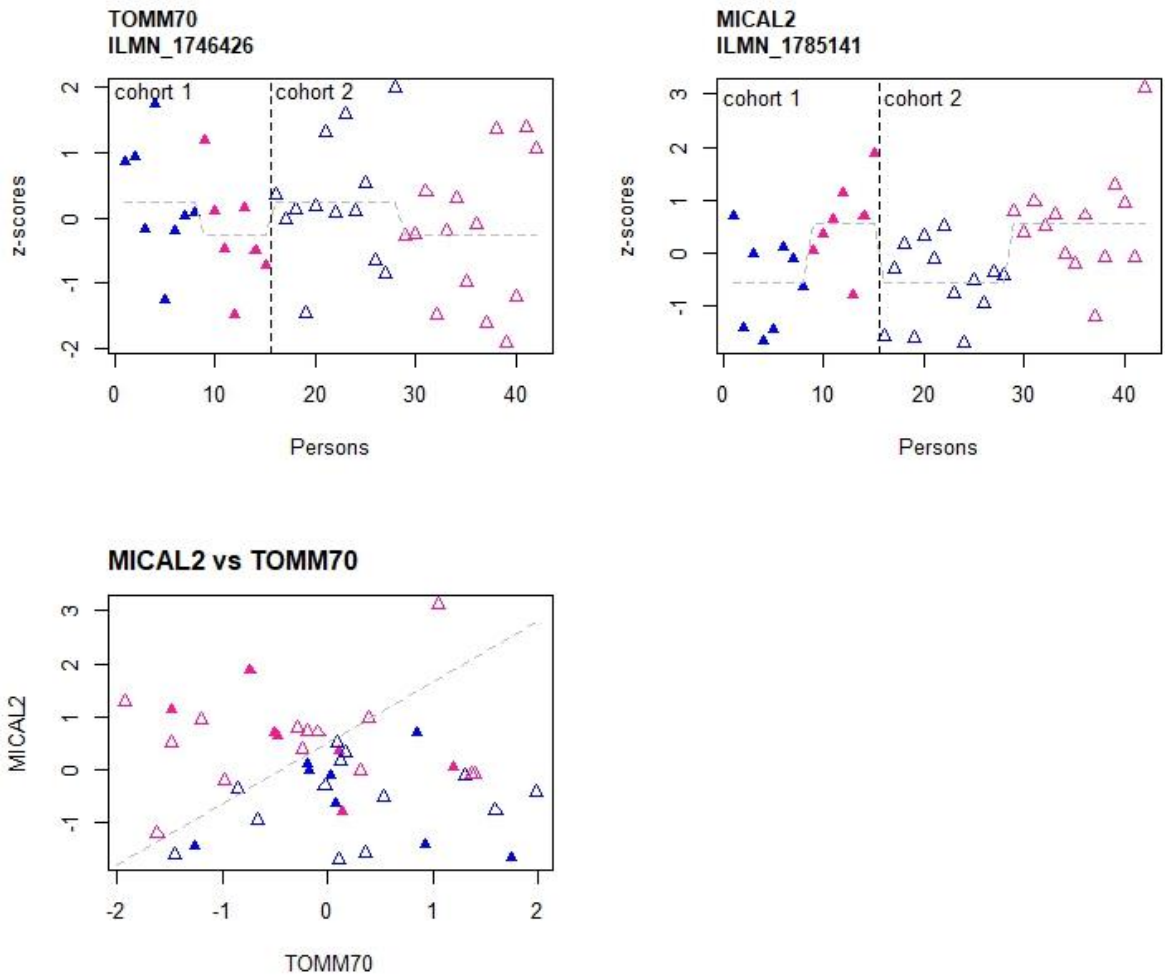

**Figure 8.** *Combined view of two and two genes.*

One of the reasons why the univariate ANOVA did not detect any significance in the gene expression between nonD and T2D is false negative. As very many univariate t-tests are applied, the p-values must be corrected for multiple comparisons, with the risk of not finding significant effects that are true. Another important problem for univariate analysis of multivariate data is that the univariate test is blind to see how different genes in combination affect the response parameter. In reality, genes and proteins etc. do not act independently of each other but instead are orchestrated in the organism. Multivariate analysis therefore enables insight into the functionality of the organisms that cannot be obtained by viewing the expression level of one feature at a time.

The present publication also shows how data from different cohorts can be combined by using GEM to increase the power of the analysis and to identify differences that are valid across cohorts.

## Competing interests

We declare no conflicting interests.